\documentclass[pra,twocolumn]{revtex4}
\usepackage{makeidx}
\usepackage{amssymb}
\usepackage{amsfonts}
\usepackage{amsmath}
\usepackage{graphicx}
\usepackage{bm}
\usepackage{braket}

\begin{document}

\title{Stable High-Order Vortices in Spin-Orbit-Coupled Spin-1 Bose-Einstein Condensates}
\author{Xin-Feng Zhang$^{1}$}
\author{Huan-Bo Luo$^{1,2,3}$}
\email{huanboluo@fosu.edu.cn}
\author{Josep Batle$^{4,5}$}
\author{Bin Liu$^{1,2}$ }
\email{binliu@fosu.edu.cn}
\author{Yongyao Li$^{1,2}$}

\affiliation{$^1$School of Physics and Optoelectronic Engineering, Foshan University, Foshan 528000, China}
\affiliation{$^2$Guangdong-Hong Kong-Macao Joint Laboratory for Intelligent Micro-Nano Optoelectronic Technology, Foshan University, Foshan 528225, China}
\affiliation{$^3$Department of Physics, South China University of Technology, Guangzhou 510640, China}
\affiliation{$^4$Departament de F\'isica UIB and Institut d'Aplicacions Computacionals de Codi Comunitari (IAC3), Campus UIB, E-07122 Palma de Mallorca, Balearic Islands, Spain}
\affiliation{$^5$CRISP -- Centre de Recerca Independent de sa Pobla, sa Pobla, E-07420, Mallorca, Spain}

\begin{abstract}
The present contribution explores phase transitions that occur in the ground state (GS) of spin-1 
Bose-Einstein condensates (BECs) with spin-orbit coupling (SOC) under the action
of gradient magnetic fields. By solving the corresponding linearized system in an exact fashion, we identify
the conditions under which the GS phase transitions occur, thus transforming excited states into GS. The study of the full nonlinear system, including both density-density and spin-spin interactions, is numerically analyzed. For the case of repulsive spin-spin interactions, the results resemble the linear case, while attractive spin-spin interactions lead to the formation of mixed-states near the GS phase-transition points. 
Additionally, higher-order vortex states are found to be stable even in the 
nonlinear regime. These findings demonstrate that arbitrary winding numbers can be achieved as 
corresponding to stable GS and thus contributing to the understanding of 
topological properties in SOC BECs.
\end{abstract}

\maketitle

\section{Introduction}
Atomic Bose-Einstein condensates (BECs), due to their remarkable experimental controllability, 
have emerged as versatile platforms for simulating and exploring a wide range of phenomena, previously 
encountered in more complex forms in condensed matter physics~\cite{RepProgPhys.75.082401,Lewenstein}. 
Of particular interest is the emulation of spin-orbit coupling (SOC) in BECs, originally known as 
the interaction between an electron's spin and its orbital motion in semiconductors. SOC plays a pivotal role 
in various quantum phenomena with diverse applications, including the spin Hall effect~\cite{RevModPhys.82.1959}, 
topological insulators~\cite{RevModPhys.82.3045}, and spintronic device design~\cite{RevModPhys.76.323}. 
During the past decade, synthetic SOC has been predicted and experimentally realized in one~\cite{nature09887},
two~\cite{nature22,Science.354.83}, and three~\cite{Juzeliunas,Science.372.271} dimensions in binary atomic gases, 
offering pristine platforms to investigating novel topological effects. The interplay between tailored 
SOC and intrinsic nonlinearities of ultracold atoms gives rise to a rich variety of new states 
characterized by nontrivial topological properties and dynamical stability. Skyrmions~\cite{PhysRevLett.109.015301}, vortices~\cite{Kawakami,Drummond,Sakaguchi,PhysRevE.89.032920,PhysRevE.94.032202,CNSNS}, 
diverse species of solitons~\cite{PhysRevLett.110.264101,1D sol 2,1D sol 3,1D sol 4,Cardoso,Lobanov,2D 
SOC gap sol Raymond,SOC 2D gap sol Hidetsugu,low-dim SOC, Subhasis}, or plane waves~\cite{prl-115-253902,pla-418-127696,fop-17-6,pra-109-053307,pre-103-022204,pra-053608,pra-033621} constitute typical examples. 
These experimental and theoretical advances have been comprehensively reviewed in Refs.~\cite{Spielman,
Galitski,Ohberg,Zhai,SOC-sol-review}.

Achieving stable high-order vortex solitons in BECs has become a difficult task~\cite{PhysRevA.108.044210}. This is due to the fact that the ground state of the system is inherently stable~\cite{pra-011607,pra-023606,prl-015301}, and thus further lowering of the energies so that it becomes the new effective ground state is clearly non-trivial. One possible way to address the situation is to induce a ground-state phase transition. 
In such scenarios, the energy gap between the GS and the first excited 
state closes at the transition point. Notably, it has recently been demonstrated that in 
one-~\cite{PhysRevA.106.063311} and two-dimensional~\cite{PhysRevA.109.013326} two-component 
SOC BECs (i.e., spin-1/2 systems), excited states with arbitrarily high quantum numbers can 
be converted into the GS by tuning the SOC strength and the gradient of an applied external magnetic field.

In this work, we extend these ideas to spin-1 (three-component) SOC BECs, 
aiming to unveil the possibility of GS phase transitions in such systems. Compared to their spin-1/2 counterparts~\cite{duan,Cui,wang} ,
spin-1 systems naturally offer a richer internal structure and more diverse physical behaviors~\cite{Ma-Jia,Adhikari}. 
For instance, depending on the sign of the spin-spin interaction, the GS of a spin-1 condensate 
can be either a polar state ~\cite{ho,pra-023629,pra-023617}(for repulsive interactions) or a ferromagnetic state (for attractive interactions)~\cite{cpb-030302,pra-061601,pra-023618,pra-023641}.
We introduce a spin-1 SOC system incorporating both a harmonic trapping potential and 
a gradient magnetic field. In the linear regime, when the SOC strength equals the 
magnetic field gradient, the system can be solved exactly. An analysis of the energy 
spectrum shows that a simultaneous increase of these parameters leads to a reduction 
in energy, with higher excited states exhibiting faster energy decay. As a result, and for a  
suitable choice of parameters, the system undergoes a GS phase transition, wherein any excited 
state can become the new GS.

We further explore the nonlinear regime, including both repulsive and attractive 
spin-spin interactions, by recourse to numerical simulations. The solutions in this case 
are mixed states, that is, essentially superpositions of multiple linear eigenstates. 
Notably, in the presence of a strong attractive spin-spin interaction, the system gives rise to mixed states, 
which spontaneously shift from the trap center to the edge. This behavior is supported by 
analytical considerations. Additionally, we analyze the corresponding nonlinear states by 
tracking the magnetization of the eigenstates. The present contribution is organized as follows: 
in Section~II we introduce the model and its description.  Section~III presents the exact linear solution. 
In section~IV we provide numerical results for the nonlinear system with repulsive and attractive 
interactions. The stability of the vortex and mixed states against small
perturbations is discussed in Sec.~V. Finally, some conclusions are drawn in Section~VI.

\section{The model}
We consider a 2D spin-1 BEC with SOC~\cite{njop-063037}, confined within a harmonic potential 
and subjected to a gradient magnetic field. In its dimensionless form, the 
harmonic potential well is given by $V = (x^2 + y^2)/2$, and the 
gradient magnetic field is $\mathbf{B} = (-\alpha x, -\alpha y, -1)$, 
where $-\alpha$ corresponds to the magnetic field gradient along the $x$ 
and $y$ directions, and $B_z=-1$ indicates a uniform bias magnetic field of one 
unit along the $-z$ direction. Additionally, the SOC term is given by 
$H_{\text{SOC}} =\beta(F_xp_y-F_yp_x)= i\beta (F_y\partial_x-F_x\partial_y)$, 
where $\beta$ denotes the spin-orbit coupling strength~\cite{prl-035302}. The vector matrix 
$\mathbf{F}=(F_x,F_y,F_z)$ are the ``Pauli matrices" for the spin-1 angular 
momentum representation~\cite{pra-043019}, and can be expressed as
\begin{equation}
	\begin{aligned} & F_x=\frac{1}{\sqrt{2}}
		\begin{bmatrix}0 & 1 & 0\\ 
			1 & 0 & 1 \\ 
			0 & 1 & 0 
		\end{bmatrix}, \quad
		F_y=\frac{1}{\sqrt{2}} 
		\begin{bmatrix}
			 0 & -i & 0 \\ 
			 i & 0 & -i \\
			0 & i & 0 
		\end{bmatrix}, \\ 
		& F_z=
		\begin{bmatrix} 
			1 & 0 & 0\\ 
			0 & 0 & 0 \\ 
			0 & 0 & -1 
		\end{bmatrix},
	\end{aligned}  \label{F}
\end{equation}
that satisfy the usual angular momentum commutation relations. 
The dynamics of the system's spinor wave function $\Psi = (\Psi_1, \Psi_0, \Psi_{-1})^T$ 
is governed by the Gross-Pitaevskii (GP) equation~\cite{prl-035302}:  
\begin{equation}
	\begin{split}
		i\partial _{t}\Psi =& \left[ \frac{1}{2}\left(-\nabla^2+r^2\right)
		-\alpha \left(xF_{x}+yF_y\right)- F_{z} \right. \\
		& +i\beta\left(\partial_{x}F_{y}-\partial_{y}F_x\right)+c_{0}\rho +c_{2}\rho \mathbf{S}\mathbf{\cdot }\mathbf{F}\bigg]\Psi,
	\end{split}
	\label{main}
\end{equation}
where the particle number density is given by  $\rho = \Psi^\dagger \Psi$. The 
coefficients $c_0$ and $c_2$ represent the strength of the density and spin 
interactions, respectively. The spin vector is given by $\mathbf{S} = \Psi^\dagger% 
\mathbf{F} \Psi/\rho$.

The stationary-state solution of Eq.~\eqref{main} with chemical potential $\mu$ is given by~\cite{PhysRevE.94.032202}  
\begin{equation}
		\Psi(x,y,t) = \exp(-i\mu t) \psi(x,y),
	\label{stationary}
\end{equation}
with the stationary wave function $\psi = (\psi_1, \psi_0, \psi_{-1})^T$ satisfying the following 
equation:
\begin{equation}
	\begin{split}
		\mu\psi =& \left[ \frac{1}{2}\left(-\nabla^2+r^2\right)
		-\alpha \left(xF_{x}+yF_y\right)- F_{z} \right. \\
		& +i\beta\left(\partial_{x}F_{y}-\partial_{y}F_x\right)+c_{0}\rho +c_{2}\rho \mathbf{S}\mathbf{\cdot }\mathbf{F}\bigg]\psi.
	\end{split}
	\label{stationary_EQ}
\end{equation}

Equations~(2) and (4) are written in dimensionless form. In physical units, we consider the 
condensate to be composed of $^{87}\mathrm{Rb}$ atoms confined in a harmonic trap with frequency 
$\omega = 10\,\mathrm{Hz}$. The number of atoms in the
condensates is $1000$, which is sufficient for the experimental observation
of the predicted patterns in full detail. The characteristic length, time
and energy are identified as $l=\sqrt{\hbar /m_{\mathrm{at}}\omega }=8.55$ $%
\mathrm{\mu }$m, $\tau =1/\omega =100$ ms, and $\epsilon =\hbar \omega
=1.05\times 10^{-33}$ J, where $m_{\mathrm{at}}=1.44\times 10^{-25}$ kg is
the atomic mass of $^{87}$Rb. The strength of SOC, denoted by 
$\beta=l\pi/\left(\sqrt{3}\lambda\right)$, where $\lambda$ represents the 
wavelength of the laser, can be adjusted across a wide range depending on the specific 
configurations of the laser system \cite{BEC-SOC GP eqns}. Moreover, the shorter the wavelength of the 
laser, the greater the SOC strength. For instance, the Nd:YAG lasers typically 
emit light with a wavelength of 1064 nm, corresponding to a SOC strength of 
$\beta=1.47$, while the He-Ne lasers emit light with a wavelength of 633 nm, 
resulting in a higher SOC strength of $\beta=2.45$. The gradient magnetic field 
can be experimentally implemented using a pair of Helmholtz coils. By adjusting 
the current and the number of turns in the coils, the gradient field $\alpha$ can be 
independently and flexibly controlled.

In the following, we will use analytical and numerical methods to seek the static solutions of 
Eq.~\eqref{stationary_EQ} in the linear and nonlinear cases, respectively.

\begin{figure}[bh]
	\centering
	\includegraphics[width=3.4in]{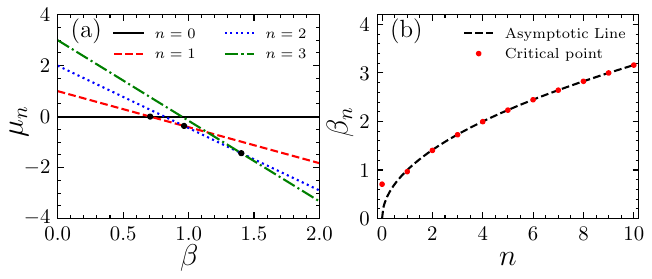}
	\caption{(a) The corresponding chemical potential $\mu_n(\beta)$ plotted
	pursuant to Eq.~\eqref{mu}. The dots are values of $\beta_n$ defined by Eq.~\eqref{beta}. 
	(b) Scatter points show the dependence of the $n$th critical point $\beta_n$ on $n$; 
	the dashed line indicates the asymptotic behavior given by Eq.~\eqref{asymptotic}.}
	\label{figure1}
\end{figure}

\section{The Exact Solution of Linear System}
First, let us solve the linearized GP Eq.~\eqref{main}, which corresponds to the case of $c_0 = c_2 = 0$. 
Moreover, when the magnetic field gradient and spin-orbit coupling are equal in strength, i.e., 
$\alpha = \beta$, the system admits an exact solution. By applying the above two simplifications 
and substituting Eq.~\eqref{stationary} into Eq.~\eqref{main}, the system can be reduced to the following eigenvalue equation:  
\begin{equation}
	H\psi = \mu\psi,
	\label{hp}
\end{equation}
where the Hamiltonian is given by
\begin{equation}
	\begin{aligned}
		H= &\frac{1}{2}\left(-\nabla^2+r^2\right)
		-\beta\left[\left(y-i\partial_{x}\right)F_{y}+\left(x+i\partial_{y}\right)F_x\right]- F_{z}\\
		&=\begin{bmatrix}
			\hat{a}^\dagger\hat{a}+\hat{b}^\dagger\hat{b}&-\sqrt{2}\beta\hat{a}^\dagger&0 \\	
			-\sqrt{2}\beta\hat{a}&\hat{a}^\dagger\hat{a}+\hat{b}^\dagger\hat{b}+1&-\sqrt{2}\beta\hat{a}^\dagger\\
			0&-\sqrt{2}\beta\hat{a}&\hat{a}^\dagger\hat{a}+\hat{b}^\dagger\hat{b}+2\\
		\end{bmatrix},
	\end{aligned}
		\label{H}
\end{equation}
where the creation and annihilation operators for Landau-level are introduced as
\begin{equation}
	\begin{aligned}
		&\hat{a}^\dagger=\frac{(x-iy-\partial_x+i\partial_y)}{2},
		\hat{a}=\frac{x+iy+\partial_x+i\partial_y}{2},\\
		&\hat{b}^\dagger=\frac{(x+iy-\partial_x-i\partial_y)}{2},
		\hat{b}=\frac{x-iy+\partial_x-i\partial_y}{2}.
		\label{ab}
	\end{aligned}		
\end{equation}
Except for $[ \hat{a}, \hat{a}^\dagger ] = 1$ and $[ \hat{b}, \hat{b}^\dagger ] = 1$, 
all other operators commute with each other. This means that $\hat{a}, \hat{a}^\dagger$ 
and $\hat{b}, \hat{b}^\dagger$ each form an independent harmonic oscillator algebra.

In the following, and for later convenience, we introduce the series form of the Landau-level wave function as~\cite{landau}:
\begin{equation}
	\begin{split}
	& f_{n,m}(r,\theta ) \\
	& =\frac{\exp \left( im\theta -r^{2}/2\right) }{\sqrt{\pi n!(n+m)!}}%
	\sum_{k=0}^{n+m}C_{n+m}^{k}A_{n}^{k}(-1)^{k}r^{2(n-k)+m},
	\end{split}
	\label{Landau}
\end{equation}%
where $m $ represents the winding number (also referred to as vorticity or the magnetic quantum number), 
while $n $ is an auxiliary quantum number. The Landau-level index is given by $n + m $. The quantum 
numbers go as $n = 0,1,2,\dots $ and $m = -n, -n+1, -n+2, \dots $. The binomial coefficients are 
defined as $C_{n}^{m} = \frac{n!}{m!(n-m)!} $, and the coefficients $A_{n}^{m} $ are given by
 $A_{n}^{m} = \frac{n!}{(n-m)!} $ for $m \leq n $, while $A_{n}^{m} \equiv 0 $ for $m > n $.  
The action of the creation and annihilation operators $\hat{a}^{\dagger} $  and $\hat{b}^{\dagger} $
on the wave function (\ref{Landau}) is expressed as  
\begin{equation}
	\begin{array}{ll}
		\hat{a}^{\dagger } f_{n,m} = \sqrt{n+1} f_{n+1,m-1}, & \hat{a} f_{n,m} = \sqrt{n} f_{n-1,m+1}, \\  
		\hat{b}^{\dagger } f_{n,m} = \sqrt{n+m+1} f_{n,m+1}, & \hat{b} f_{n,m} = \sqrt{n+m} f_{n,m-1}.
	\end{array}\label{rise_down}
\end{equation}

\begin{figure}[tbh]
	\centering
	\includegraphics[width=3.4in]{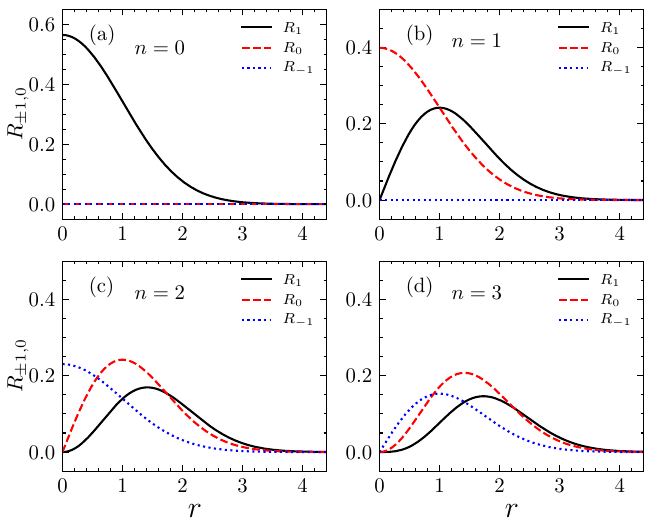}
	\caption{Profiles of the radial wavefunctions $R_{\pm1,0}^{(n)}$, defined as per Eqs.~\eqref{R} and~\eqref{R0}, 
	with quantum numbers (a) $n=0$, (b) $n=1$, (c) $n=2$ and (d) $n=3$.}
	\label{figure2}
\end{figure}

With the aid of Eqs.~\eqref{hp},~\eqref{H} and~\eqref{rise_down}, the exact solution of the eigenvalue equation (4) can be easily obtained as
\begin{equation}
	\psi=\frac{1}{\sqrt{4n-2}}\begin{pmatrix}
		\sqrt{n}f_{n,m}(r,\theta)\\
		\sqrt{2n-1}f_{n-1,m+1}(r,\theta)\\
		\sqrt{n-1}f_{n-2,m+2}(r,\theta)
	\end{pmatrix},n=1,2,3,\cdots
	\label{psii}
\end{equation}
whilst for $n=0$, we have
\begin{equation}
	\psi=\begin{pmatrix}
		f_{0,m}(r,\theta)\\
		0\\
		0
	\end{pmatrix}.
	\label{psii0}
\end{equation}
The corresponding chemical potential is 
\begin{equation}
	\mu = \left\{\begin{array}{ll}
		m,&n=0\\
		2n+m-\beta\sqrt{4n-2},&n=1,2,3,\cdots\\
	\end{array}\right.
\end{equation}
Since we are only concerned with the ground state of the system, and considering the range of the quantum number $m$, the chemical potential is minimized when $m = -n$, and it can be simplified in the form 
\begin{equation}
	\mu_n(\beta) = \left\{\begin{array}{ll}
		0,&n=0\\
		n-\beta\sqrt{4n-2},&n=1,2,3,\cdots\\
	\end{array}\right.
	\label{mu}
\end{equation}
The curve of the chemical potential $\mu_n(\beta)$ as a function of $\beta$ is shown in Fig.~\ref{figure1}.
As shown in Fig.~\ref{figure1}, spin-orbit coupling and the gradient magnetic field reduce the system's energy, 
and the higher the energy level (the larger $n$), the greater the energy reduction. Therefore, as long as a 
suitable $\beta$ is given, the wave function corresponding to any quantum number $n$ can become the ground state, meaning a ground-state phase transition occurs. In addition, the $n$th phase transition point 
$\beta_n$ can be determined from $\mu_n(\beta_n) = \mu_{n+1}(\beta_n)$, yielding
\begin{equation}
	\beta_n = \left\{\begin{array}{ll}
		\sqrt{2}/2,&n=0\\
		\left(\sqrt{4n+2}+\sqrt{4n-2}\right)/4,&n=1,2,3,\cdots\\
	\end{array}\right.
	\label{beta}
\end{equation}
The first four phase transition points are $\beta_0 = 0.7071$, $\beta_1 = 0.9659$, $\beta_2 = 1.4029$ and 
$\beta_3 = 1.7260$, which are represented by black dots in Fig.~\ref{figure1}(a). As $n$ increases, the critical 
points exhibit an asymptotic behavior that can be approximated by 
\begin{equation}
\beta_n \approx \sqrt{n},
	\label{asymptotic}
\end{equation}
as shown in Fig.~\ref{figure1}(b).

According to Eqs.~\eqref{Landau},~\eqref{psii} and~\eqref{psii0}, the corresponding wave 
function form shall be simplified as
\begin{equation}
	\psi=\begin{pmatrix}
		R^{(n)}_1(r)e^{-in\theta}\\
		R^{(n)}_0(r)e^{-i(n-1)\theta}\\
		R^{(n)}_{-1}e^{-i(n-2)\theta}
	\end{pmatrix},
	\label{psis}
\end{equation}
where $R_1^{(n)}(r)$, $R_0^{(n)}(r)$ and $R_{-1}^{(n)}(r)$ are the radial wave functions, which (for $n\geq1$) can be expressed as
\begin{equation}
	\begin{aligned}
		&R_1^{(n)}=\frac{1}{\sqrt{\pi(4n-2)(n-1)!}}r^{n}\exp\left(-\frac{r^2}{2}\right),\\
		&R_0^{(n)}=\frac{1}{\sqrt{2\pi(n-1)!}}r^{n-1}\exp\left(-\frac{r^2}{2}\right),\\
		&R_{-1}^{(n)}=\frac{n-1}{\sqrt{\pi(4n-2)(n-1)!}}r^{n-2}\exp\left(-\frac{r^2}{2}\right),\\
		\label{R}
	\end{aligned}		
\end{equation}
and for $n=0$
\begin{equation}
		R_1^{(0)}=\frac{1}{\sqrt{\pi}}\exp\left(-\frac{r^2}{2}\right),\quad R^{(0)}_{0}=R^{(0)}_{-1}=0.
		\label{R0}
\end{equation}
Typical profiles of the radial wave functions are plotted in Fig.~\ref{figure2}.

We observe that the wave function with quantum number $ n $ exhibits vortices with winding numbers 
$(n, n-1, n-2)$ in the three different spin components. In the nonlinear regime, vortices with higher 
winding numbers are typically unstable~\cite{pra-033319}. However, in our system, by tuning $\beta$, any quantum number 
$ n $ can transition to the ground state, enabling the realization of stable high-order vortex solitons.

\begin{figure}[tbh]
	\centering
	\includegraphics[width=3.4in]{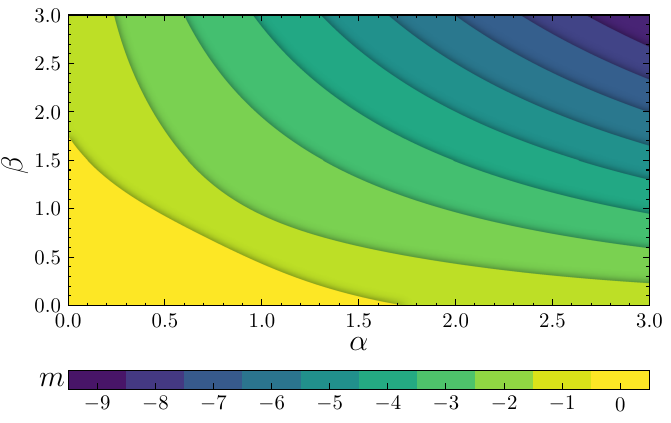}
	\caption{ The map of values of the winding number (magnetic quantum number) $m$ corresponding to GS of the linear system in the ($\alpha$, $\beta$) parameter plane.}
	\label{figure3}
\end{figure}

The above discussion is based on the special case of $\alpha = \beta$. However, in experiments, it is quite 
difficult to make these two parameters exactly equal. Therefore, we now consider a more general situation 
where $\alpha \neq \beta$, in which case $\alpha$ and $\beta$ are treated as two independent variables.
In this case, the solution of the linear system with winding number $m$ can still be expanded in terms of Landau level wavefunctions:
\begin{equation}
\begin{split}
\psi _{1}& =\sum_{n=0}^{N_t}k_{n}f_{n,m}, \\
\psi _{0}& =\sum_{n=0}^{N_t}d_{n}f_{n-1,m+1},\\
\psi _{-1}& =\sum_{n=0}^{N_t}e_{n}f_{n-2,m+2},
\end{split}
\label{general}
\end{equation}%
where $k_n$, $d_n$ and $e_n$ are coefficients to be determined. $N_t$ is the truncation order. Here, we 
choose $N_t = 50$, which is already sufficient to produce accurate results.
Substituting the solution~\eqref{general} into the linearized stationary equation~\eqref{stationary_EQ}, 
we derive a set of coupled linear equations for $k_n$, $d_n$, and $e_n$:
\begin{equation}
\begin{split}
\mu k_{n}& =\left( 2n+m\right) k_{n}-\frac{\alpha +\beta}{\sqrt{2}}\sqrt{n}%
d_{n} \\
& +\frac{\beta -\alpha}{2}\sqrt{n+m+1}d_{n+1}, \\
\mu d_{n}& =\left( 2n+m\right) d_{n}-\frac{\alpha +\beta}{\sqrt{2}}\sqrt{n}%
k_{n} \\
& +\frac{\beta -\alpha}{2}\sqrt{n+m}k_{n-1}-\frac{\alpha +\beta}{\sqrt{2}}\sqrt{n-1}e_{n}\\
&+\frac{\beta -\alpha}{2}\sqrt{n+m+1}e_{n+1},\\
\mu e_{n}& =\left( 2n+m\right) e_{n}-\frac{\alpha +\beta}{\sqrt{2}}\sqrt{n-1}%
d_{n} \\
& +\frac{\beta -\alpha}{2}\sqrt{n+m}d_{n-1}. \\
\end{split}
\label{couple}
\end{equation}%
Once the quantum number $m$ is fixed, Eq.~\eqref{couple} can be solved through numerical diagonalization of the 
corresponding matrix. By comparing the chemical potentials $\mu$ for different winding numbers $m$, we can identify the 
system's ground state.

\begin{figure}[tbh]
	\centering
	\includegraphics[width=3.4in]{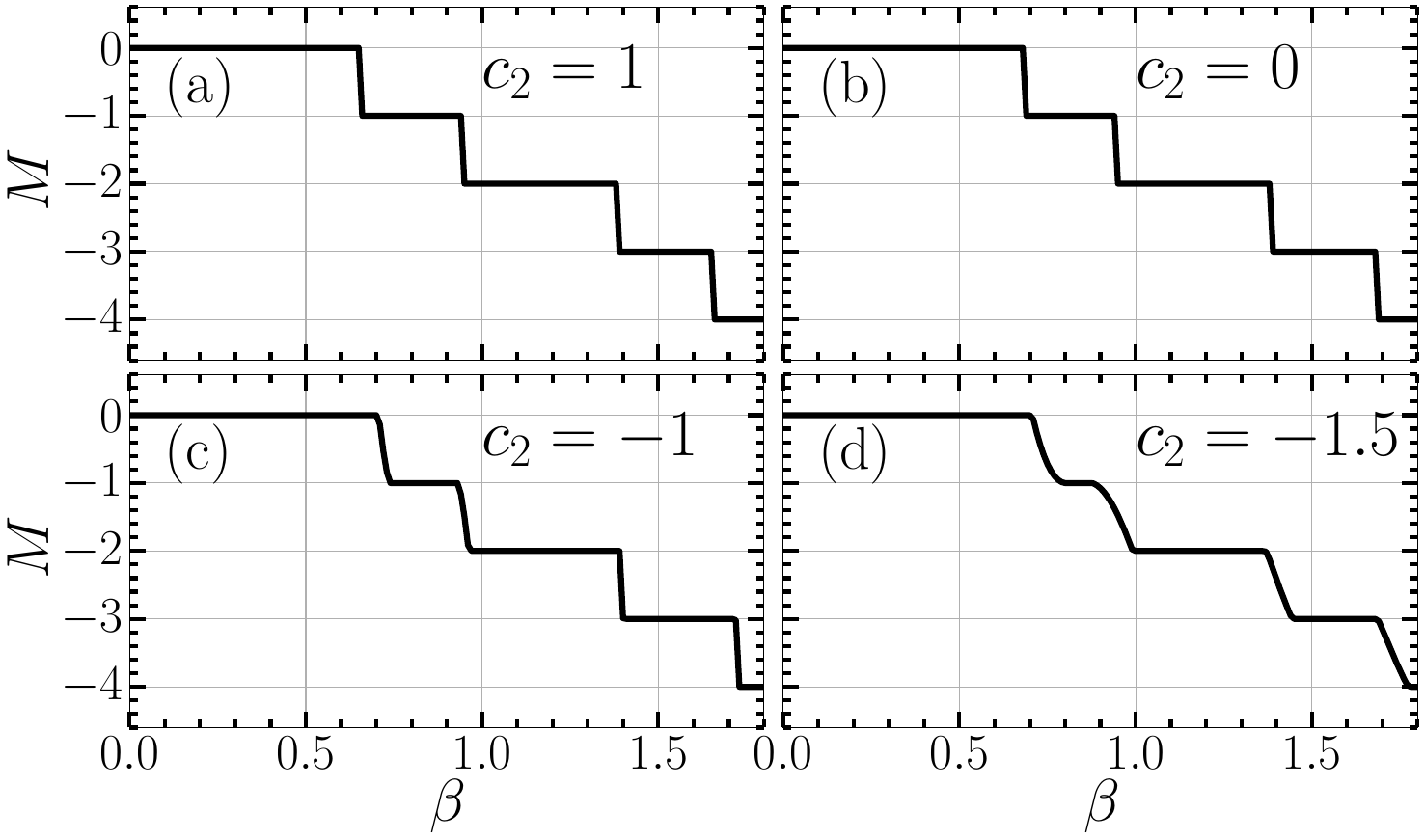}
	\caption{(a-d) Angular momentum $M$, defined as per Eqs.~\eqref{M}, as produced by the imaginary-time
	simulations of the full (nonlinear) system (\protect\ref{main}), for $\protect\beta $ varying from $0$
	to $1.8$. Other parameter in Eq.~\eqref{main} is $\protect c_0 =1$.}
	\label{figure4}
\end{figure}

The resulting map of quantum numbers $m$ corresponding to the GS in the $(\alpha, \beta)$ parameter plane, i.e., the GS 
phase diagram, is shown in Fig.~\ref{figure3}. Along the diagonal line $\alpha = \beta$, the GS predicted by this diagram agrees with
the exact result given by Eq. (20), reflecting the system's symmetry about this diagonal. The phase diagram also reveals
that GS transitions between adjacent winding numbers $m$ and $m + 1$ occur precisely at $\alpha \neq \beta$.
These results indicate that ground-state phase transitions still occur even when $\alpha \ne \beta$, which significantly
reduces the experimental difficulty of precisely matching the spin-orbit coupling strength $\beta$ and the magnetic field 
gradient $\alpha$.

\begin{figure}[tbh]
	\centering
	\includegraphics[width=3.4in]{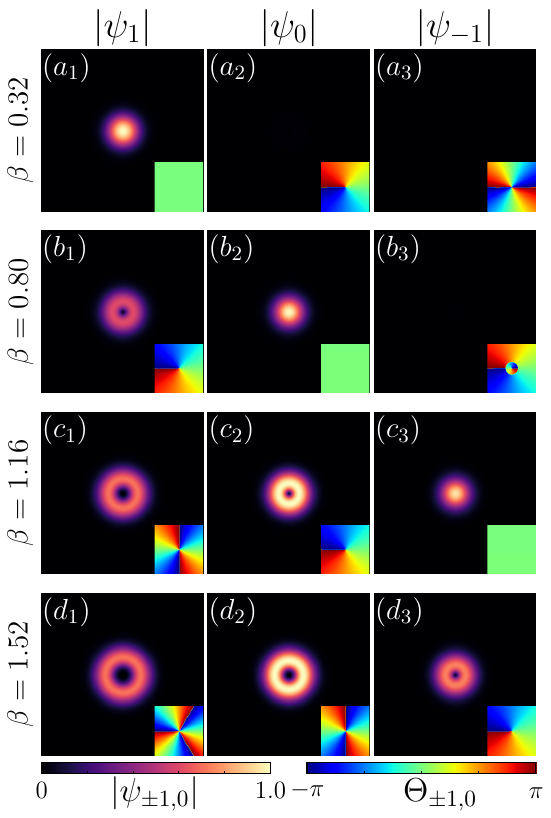}
	\caption{Distributions of the absolute values, $\left\vert \protect%
	\psi_{\pm1,0}\right\vert $, and phases, $\Theta_{\pm1,0}$, of wave functions of the
	three components in the GS of the vortex type, for $\protect\beta =0.32,\mu=0.2812$
	(panels $a_{1}-a_{3}$); $\protect\beta =0.80,\mu=0.2794$ (panels $%
	b_{1}-b_{3}$); $\protect\beta =1.16,\mu=-0.7114$
	(panels $c_{1}-c_{3}$) and  $\protect\beta =1.52,\mu=-1.7008$ (panels $%
	d_{1}-d_{3}$). Other parameters in
	Eq.~\eqref{main} are $\protect c_0=c_2 =1$.}
	\label{figure5}
\end{figure}

Since exact solutions can provide theoretical predictions, we continue to adopt the condition $\alpha = \beta$ in the 
following discussion of the nonlinear case.

\section{The Nonlinear System and its Numerical Solution}
In this section, we incorporate nonlinearity into the system, including both density 
and spin interactions~\cite{pra-023641,pra-023612}. To simplify the analysis, we fix the density interaction strength 
at $ c_0 = 1 $ and vary the spin interaction strength $ c_2 $ relative to that value. 
A positive $ c_2 $ corresponds to repulsive spin interactions, while a negative $ c_2 $ 
represents attractive spin interactions~\cite{pra-023618}. 

In order to solve the nonlinear system, we use the imaginary-time 
evolution method to numerically obtain the ground state~\cite{chiofalo,Bao}. During the 
iterative solving process, the total particle number is conserved, i.e.,  
\begin{equation}
N = \int_{-\infty}^{\infty} \int_{-\infty}^{\infty} \psi^\dagger \psi dxdy = 1.
\end{equation}

\begin{figure}[tbh]
	\centering
	\includegraphics[width=3.4in]{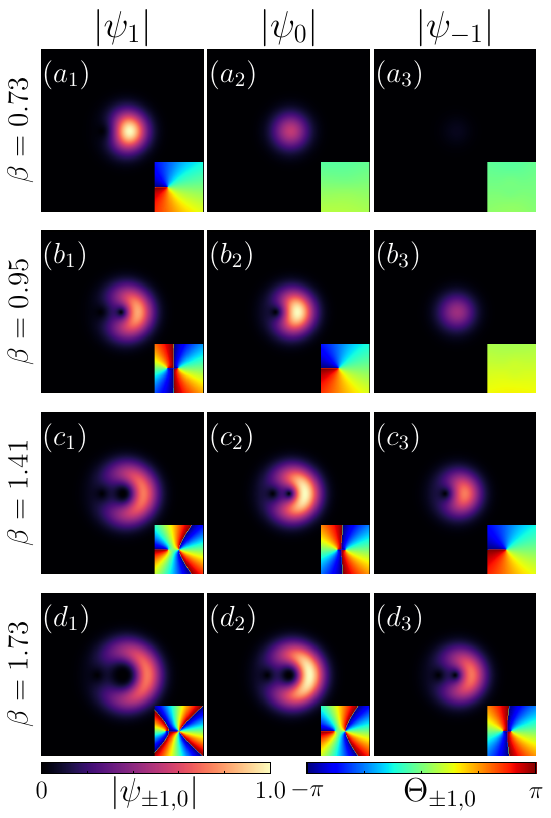}
	\caption{Distributions of the modulus, $\left\vert \protect%
	\psi_{\pm1,0}\right\vert $, and phases, $\Theta_{\pm1,0}$, of wave functions of the
	two components in the GS of the mixed-state type for $\protect\beta =0.73,\mu=-0.0857$
	(panels $a_{1}-a_{4}$), $\protect\beta =0.95,\mu=-0.3763$ (panels $b_{1}-b_{4}$), $%
	\protect\beta =1.41,\mu=-1.4979$ (panels $c_{1}-c_{4}$), $\protect\beta =1.73,\mu=-2.5081$ 
	(panels $d_{1}-d_{4}$), and $c_0 =1$, $\protect%
	c_2 =-1.5$ in Eq.~(\protect\ref{main}).}
	\label{figure6}
\end{figure}

To characterize each stationary state, we define the total angular momentum operator as 
the sum of the orbital and spin angular momentum operators~\cite{njop-043019}, namely:
\begin{equation}
	\hat{O} = L+ F_z-I,\quad M=\langle \hat{O} \rangle,
	\label{M}
\end{equation}
where $L = i(y \partial_x - x \partial_y) \equiv -i \partial_\theta$ is the orbital 
angular momentum operator, and $I$ is the $3 \times 3$ identity matrix, introduced to 
shift the quantum numbers such that the largest quantum number is set to zero.
By applying the operator $\hat{O}$ to the stationary states described by Eqs.~\eqref{psis},
~\eqref{R} and~\eqref{R0}, we can easily obtain the expectation value $M= \langle \hat{O} \rangle= -n$.
The integer values of $M$ indicate vortex states, while noninteger values of $M$ 
indicate mixed states. The dependence of $M$ on $\beta$, produced by the
numerical solution, is shown in Fig.~\ref{figure4}.

\begin{figure}[tbh]
	\centering
	\includegraphics[width=3.4in]{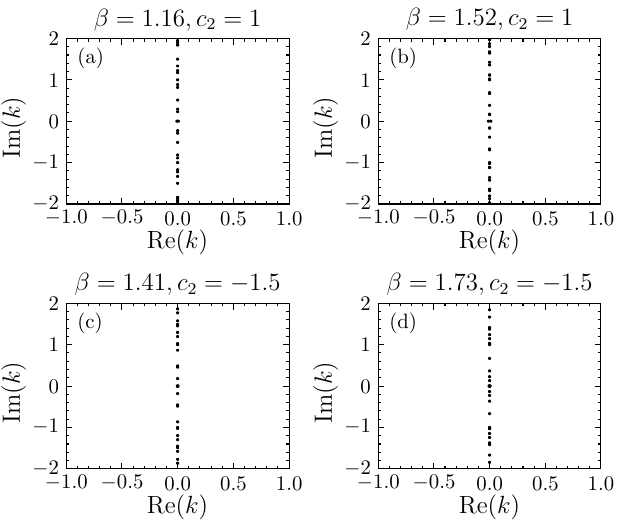}
	\caption{ The linear stability spectra of the GS solutions with
	 (a) $\beta=1.16,c_2=1$, (b) $\beta=1.52,c_2=1$, (c) $\beta=1.41,c_2=1.5$, and (d) $\beta=1.73,c_2=-1.5$.}
	\label{figure7}
\end{figure}

For different values of $c_2$, the dependencies $M(\beta)$ shown in Figs.~\eqref{figure4}(a)-(d) 
exhibit similar overall patterns, except near the phase transition points. Recall that the ground-state
phase transitions in the exact solution of the linear system are given by Eq.~\eqref{beta}—specifically, $\beta_0 = 0.
7071$, $\beta_1 = 0.9659$ and $\beta_2 = 1.4029$. As $\beta$ increases, $M$ follows a descending, staircase-like pattern,
dropping by 1 at each phase transition. The flat segments of the $M(\beta)$ curves correspond to vortex states 
(see typical examples in Fig.~\ref{figure4}), while the sloped segments near transitions correspond to mixed states 
(see examples in Fig.~\ref{figure5}). The parameter $c_2$ influences the width of the phase transition regions, which 
becomes broader for smaller values of $c_2$. This is consistent with the general trend of enhanced stabilization
of mixed states and destabilization of vortices as $c_2$ decreases.

\begin{figure}[tbh]
	\centering
	\includegraphics[width=3.4in]{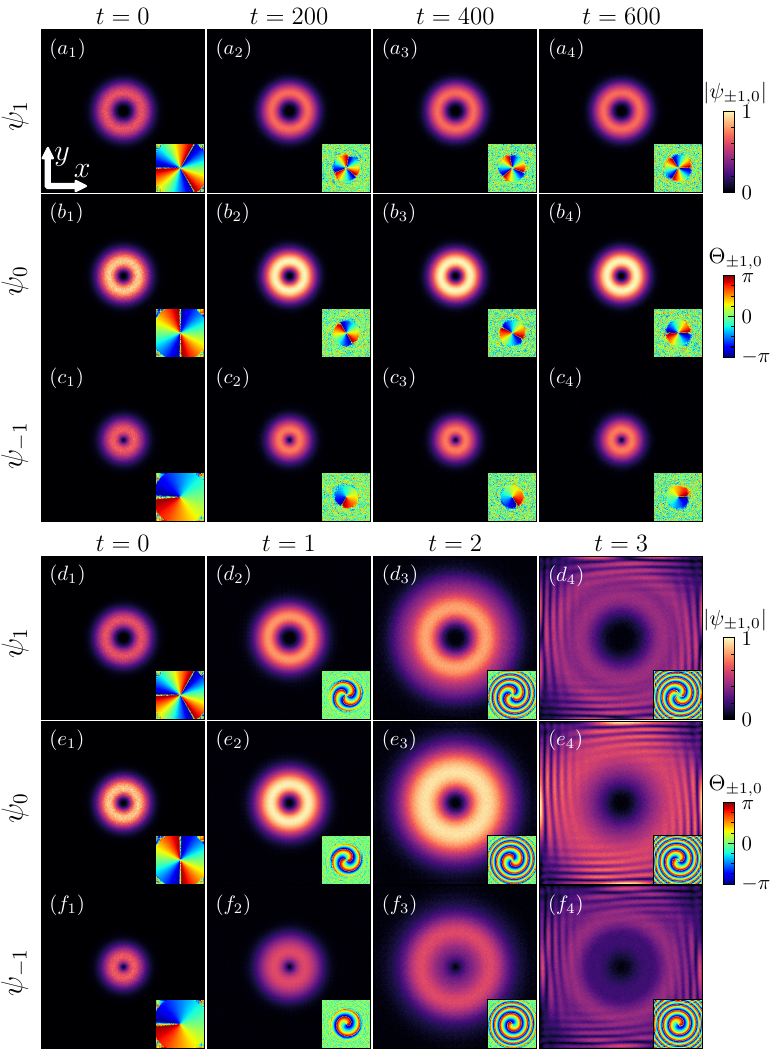}
	\caption{Numerical evolution of the perturbed ground state with parameters $\beta=1.52$ and $c_2=1$. Panels ($a_1-a_4$), ($b_1-b_4$) and ($c_1-c_4$) show the evolution governed by Eq.~\eqref{main}, which includes the trapping potential. In contrast, panels ($d_1-d_4$), ($e_1-e_4$) and ($f_1-f_4$) display the dynamics after the trap is switched off by removing the corresponding term from Eq.~(4). In each figure the bottom right corner depicts the phase distribution of the wave function.}
	\label{figure8}
\end{figure}

First, we consider the case of repulsive spin interaction with $ c_2 = 1 $.  
The corresponding numerical results are presented in Fig.~\ref{figure5}.  
For the representative values of the coupling strength $
\beta = 0.32,\; 0.80,\; 1.16$ and $ 1.52.$ The corresponding chemical potentials are  
$\mu = 0.2812,\; 0.2794,\; -0.7114$ and $-1.7008.$ The linear spectrum predictions given by Eqs.~\eqref{mu} and~\eqref{beta} indicate that the quantum number of the ground-state wave function is  
$n = 0, \; 1, \; 2$ and $3,$ respectively.

From the phase diagrams in Figs.~\ref{figure5}($a_1$)-($d_1$), we observe that the winding
numbers of the vortices are 0, 1, 2, and 3, respectively. Moreover, the modulus of 
the wave function reveals that the average radius of the vortices gradually 
increases. These results are consistent with the predictions for the linear case.

For attractive spin interactions, the situation becomes different. Near the phase 
transition point, the ground state of the system transitions into a mixed-state, 
which is a superposition of wave functions with adjacent quantum numbers~\cite{prb-094420}. For example, 
around $ \beta_n $, the ground state consists of a superposition of wave functions
with quantum numbers $ n $ and $ n+1 $. 

The numerical distributions of the wave function's modulus and phase are shown 
in Fig.~\ref{figure6}, where we set the spin interaction strength to $ c_2 = -1.5 $ 
and $ \beta = 0.73, 0.95, 1.41 $ and $ 1.73 $, which are near the linear phase 
transition points given by Eq.~\eqref{beta}. The corresponding chemical potentials are  
$\mu = -0.0857,\; -0.3763,\; -1.4979$ and $-2.5081.$ By comparing Fig.~\ref{figure6} with
Fig.~\ref{figure5}, the rotational symmetry of the modulus of these mixed-state 
wave functions is broken. According to the phase distribution, vortices with winding 
numbers 1, 2, 3, and 4 are formed in $ \psi_1 $, corresponding to Figs.~\ref{figure6}
$(a_1-d_1)$. However, the vortex singularities are not located at the center. Moreover, when 
the winding number is such that $n \geq 2 $, each vortex splits into two. The corresponding winding 
number distributions from left to right are $(1,1)$, $(1,2)$, and $(2,2)$, as shown in 
Figs.~\ref{figure6}$(b_1-d_1)$.

Based on the above discussion, we observe that in the case of repulsive spin 
interactions ($c_2 = 1$), the results closely follow the linear prediction. Thus, 
the phase transition points given by the linear analysis in Eqs.~\eqref{beta} can 
be used to generate vortex states with arbitrary winding numbers as ground states 
of the system. Even in the nonlinear regime, higher-order vortices remain stable, 
as they still correspond to the ground state.

\section{Stability Analysis of the Solutions}

In general, the ground state is expected to be stable, but for completeness we verify the stability 
of these solutions using two approaches, including linear stability analysis and real-time evolution.

To perform the linear stability analysis for the vortex and mixed states, we introduce perturbed solutions in the form
\begin{equation}
	\Psi(x,y,t)=\!e^{-i\mu t}\left[\psi(x,y)+w(x,y)e^{\lambda t}+v^*(x,y)e^{\lambda ^* t}\right],
	\label{psit}
\end{equation}
where $w=(w_1,w_{0},w_{-1})^T$ and $v=(v_1,v_0,v_{-1})^T$ are small perturbations, and $\lambda$ is the respective (complex) eigenvalue 
(the stability implying that all eigen values have $\text{Re}(\lambda)=0$). Inserting the perturbed solution in
Eq.~\eqref{main} and linearizing, one arrives at the linear eigenvalue problem:
\begin{equation}
	\begin{pmatrix}L_1&L_2 \\ L_2^* & L_1^*\end{pmatrix}\begin{pmatrix}w \\ v\end{pmatrix}=\lambda\begin{pmatrix}w \\ v\end{pmatrix}.
	\label{eg}
\end{equation}
The operators $L_1$ and $L_2$ can be written in the compact form:
\begin{equation}
\begin{aligned}
L_1 &= \frac{i\nabla^2 - i r^2}{2} 
+ \beta\left[(\partial_x - i y)F_y - (\partial_y - i x) F_x\right] + iF_z  \\
& + i\mu- i c_0(\psi^\dagger\psi + \psi\psi^\dagger) 
- i c_2(\mathbf{F}\psi\cdot\psi^\dagger\mathbf{F}^\dagger + \psi^\dagger\mathbf{F}\psi\cdot\mathbf{F}),\\
L_2&=-ic_0\psi\psi^T-ic_2\mathbf{F}\psi\cdot\psi^T\mathbf{F}^T.
\end{aligned}
\label{L1}
\end{equation}
For the sake of computational convenience, we also provide the explicit expansions of several complicated terms:
\begin{equation}
\begin{aligned}
&\mathbf{F}\psi \cdot \psi^\dagger \mathbf{F}^\dagger\\
&=\begin{pmatrix}
|\psi_1|^2 + |\psi_0|^2 & \psi_0 \psi_{-1}^* & -\psi_1 \psi_{-1}^* \\
\psi_{-1}\psi_0^* & |\psi_1|^2 + |\psi_{-1}|^2 & \psi_1 \psi_0^* \\
-\psi_{-1}\psi_1^* & \psi_0 \psi_1^* & |\psi_0|^2 + |\psi_{-1}|^2
\end{pmatrix},\\
\end{aligned}
\end{equation}
\begin{equation}
\begin{aligned}
&\psi^\dagger \mathbf{F} \psi \cdot \mathbf{F} \\
&=\begin{pmatrix}
|\psi_1|^2 - |\psi_{-1}|^2 & \psi_1 \psi_0^* + \psi_0 \psi_{-1}^* & 0 \\
\psi_1^*\psi_0 + \psi_0^*\psi_{-1} & 0 & \psi_1 \psi_0^* + \psi_0 \psi_{-1}^* \\
0 & \psi_1^*\psi_0 + \psi_0^*\psi_{-1} & -|\psi_1|^2 + |\psi_{-1}|^2
\end{pmatrix},\\
\end{aligned}
\end{equation}
\begin{equation}
\begin{aligned}
&\mathbf{F}\psi \cdot \psi^T \mathbf{F}^T\\
&=\begin{pmatrix}
\psi_1^2 & \psi_1 \psi_0 & \psi_0^2 - \psi_1 \psi_{-1} \\
\psi_1 \psi_0 & 2 \psi_1 \psi_{-1} & \psi_0 \psi_{-1} \\
\psi_0^2 - \psi_1 \psi_{-1} & \psi_0 \psi_{-1} & \psi_{-1}^2
\end{pmatrix}.
\end{aligned}
\end{equation}

By substituting the ground-state vortex and mixed states obtained in the previous section into 
Eq.~\eqref{eg}, we compute the corresponding eigenvalue spectra, as shown in Fig.~\ref{figure7}.
Figures~\ref{figure7}(a) and (b) present the linear stability spectra of the vortex states shown 
in Figs.~4(c$_{1-3}$) and (d$_{1-3}$), respectively, while Figs.~\ref{figure7}(c) and (d) correspond 
to the linear stability spectra of the mixed states in Figs.~\ref{figure6}(c$_{1-3}$) and (d$_{1-3}$). 
It is evident that the eigenvalues possess zero real parts, indicating that the ground-state solutions, 
whether vortex or mixed states, are stable.

In addition to linear stability analysis, we also examine the real-time evolution of the 
solutions under small perturbations.
First, we added white noise to the ground-state wave functions,
with the maximum noise intensity reaching $10\%$ of the wave
function's amplitude. Subsequently, real-time evolution was
applied to the perturbed wave functions. The results of numerical simulations are
presented in Figs.~\ref{figure8}(a-c), showcasing 
a vortex state with strength of SOC $\beta=1.52$, and repulsive interaction $c_2=1$. 
It can be seen that at $t = 0$,  the wave functions exhibit numerous noise points. However,
as time progresses, by $t = 200$ the noise points on the wave
functions disappear. Furthermore, in the subsequent evolution,
the phases of the wave functions rotate around the vortex
centers, while the amplitude distribution no longer undergoes
changes. The results completely verify the stability of the
ground states in all cases.

We also explored the evolution of the vortex state in the limit of vanishing trap frequency, 
as shown in Figs.~\ref{figure8}(d-f). We first prepared the initial state at time $t = 0$ with parameters 
$\beta = 1.52$ and $c_2 = 1$. Then, the trapping potential was turned off. As time evolves, 
we observe that the condensate expands away from the origin at $t = 1$ and $t = 2$. By $t = 3$, 
the condensate reaches the numerical boundary and undergoes unphysical interference due to 
reflections. This result highlights the essential role of the harmonic  
potential in maintaining the stability of the system.

\section{Conclusion}
This study investigates the ground-state (GS) phase transitions in spin-1 Bose-Einstein 
condensates (BECs) under the influence of spin-orbit coupling (SOC) and gradient magnetic 
fields. The linear system is analytically solved, revealing that the system can undergo 
GS phase transitions, allowing for the transformation of excited states into the GS by 
adjusting the system's parameters. Numerical solutions for the nonlinear system, 
considering both repulsive and attractive spin-spin interactions, are obtained, 
showing that stable high-order vortex solitons can be formed, even in the nonlinear 
regime. These vortex states remain stable when they correspond to the GS, and the 
results confirm the predictions of the linear theory. Furthermore, the system's 
response to varying parameters demonstrates that vortex states with arbitrary 
winding numbers can be regarded as stable GS, offering a deeper understanding 
of topological phenomena in spin-orbit coupled systems.

\section*{Acknowledgments}

This work was supported by the Guangdong Basic and Applied Basic Research Foundation 2023A1515110198, 2024A1515030131, 2025A1515011128,
National Natural Science Foundation of China through grants Nos. 12274077,
12475014 and 62405054, the Research Found of the Guangdong-Hong Kong-Macao
Joint
Laboratory for Intelligent Micro-Nano Optoelectronic Technology through
grant No. 2020B1212030010.


\begin{thebibliography}{999}
\bibitem{RepProgPhys.75.082401} P. Hauke, F. M. Cucchietti, L. Tagliacozzo,
I. Deutsch, and M. Lewenstein, Can one trust quantum simulators? Rep. Prog.
Phys. \textbf{75}, 082401 (2012).

\bibitem{Lewenstein} M. Lewenstein, A. Sanpera, and V. Ahufinger, \textit{%
	Ultracold Atoms in Optical Lattices: Simulating Quantum Many-Body Systems}
(Oxford: Oxford University Press, 2012).

\bibitem{RevModPhys.82.1959} D. Xiao, M.-C. Chang, and Q. Niu, Berry phase
effects on electronic properties, \rmp\textbf{82}, 1959 (2010).

\bibitem{RevModPhys.82.3045} M. Z. Hasan and C. L. Kane, Colloquium:
Topological insulators, \rmp\textbf{82}, 3045 (2010).

\bibitem{RevModPhys.76.323} I. \u{Z}uti\'{c}, J. Fabian, and S. D. Sarma,
Spintronics: Fundamentals and applications, \rmp\textbf{76}, 323 (2004).

\bibitem{nature09887} Y.-J. Lin, K. Jim\'{e}nez-Garc\'{\i}a, and I. B.
Spielman, Spin-orbit-coupled Bose-Einstein condensates, \nat\textbf{471}, 83
(2011).

\bibitem{nature22} L. Huang, Z. Meng, P. Wang, P. Peng, S.-L. Zhang, L. Chen, D. Li, Q. Zhou, and J. Zhang, Experimental realization of two-dimensional synthetic spin-orbit coupling in ultracold Fermi gases, Nat. Phys. \textbf{12}, 540 (2016).


\bibitem{Science.354.83} Z. Wu, L. Zhang, W. Sun, X.-T. Xu, B.-Z. Wang,
S.-C. Ji, Y. Deng, S. Chen, X.-J. Liu, and J.-W. Pan, Realization of
two-dimensional spin-orbit coupling for Bose-Einstein condensates, Science
\textbf{354}, 83 (2016).

\bibitem{Juzeliunas} B. M. Anderson, G. Juzeli\={u}nas, V. M. Galitski, and
I. B. Spieman, Synthetic 3D spin-orbit coupling, \prl\textbf{108}, 235301
(2012).

\bibitem{Science.372.271} Z.-Y. Wang, X.-C. Cheng, B.-Z. Wang, J.-Y. Zhang,
Y.-H. Lu, C.-R. Yi, S. Niu, Y. Deng, X.-J. Liu, S. Chen, and J.-W. Pan,
Realization of an ideal Weyl semimetal band in a quantum gas with 3D
spin-orbit coupling, Science \textbf{372}, 271 (2021).

\bibitem{PhysRevLett.109.015301} T. Kawakami, T. Mizushima, M. Nitta, and K.
Machida, Stable Skyrmions in SU(2) Gauged Bose-Einstein Condensates, \prl%
\textbf{109}, 015301 (2012).

\bibitem{Kawakami} T. Kawakami, T. Mizushima and K. Machida, Textures of F=2
spinor Bose-Einstein condensates with spin-orbit coupling, Phys. Rev. A
\textbf{84}, 011607 (2011).

\bibitem{Drummond} B. Ramachandhran, B. Opanchuk, X.-J. Liu, H. Pu, P. D.
Drummond, and H. Hu, Half-quantum vortex state in a spin-orbit-coupled
Bose-Einstein condensate, \pra\textbf{85}, 023606 (2012).

\bibitem{Sakaguchi} H. Sakaguchi and B. Li, Vortex lattice solutions to the
Gross-Pitaevskii equation with spin-orbit coupling in optical lattices, \pra%
\textbf{87}, 015602 (2013).

\bibitem{PhysRevE.89.032920} H. Sakaguchi, B. Li, and B. A. Malomed,
Creation of two-dimensional composite solitons in spin-orbit-coupled
self-attractive Bose-Einstein condensates in free space, \pre\textbf{89},
032920 (2014).

\bibitem{PhysRevE.94.032202} H. Sakaguchi, E. Y. Sherman, and B. A. Malomed,
Vortex solitons in two-dimensional spin-orbit coupled Bose-Einstein
condensates: Effects of the Rashba-Dresselhaus coupling and Zeeman
splitting, \pre\textbf{94}, 032202 (2016).

\bibitem{CNSNS} H.-B. Luo, B. A. Malomed, W.-M. Liu, and L. Li, Bessel
vortices in spin-orbit-coupled binary Bose-Einstein condensates with Zeeman
splitting, Communications in Nonlinear Science and Numerical Simulation,
\textbf{115}, 106769 (2022).

\bibitem{PhysRevLett.110.264101} V. Achilleos, D. J. Frantzeskakis, P. G.
Kevrekidis, and D. E. Pelinovsky, Matter-Wave Bright Solitons in Spin-Orbit
Coupled Bose-Einstein Condensates, \prl\textbf{110}, 264101 (2013).

\bibitem{1D sol 2} Y. Xu, Y. Zhang, and B. Wu, Bright solitons in
spin-orbit-coupled Bose-Einstein condensates, \pra\textbf{87}, 013614 (2013).

\bibitem{1D sol 3} L. Salasnich and B. A. Malomed, Localized modes in dense
repulsive and attractive Bose-Einstein condensates with spin-orbit and Rabi
couplings, \pra\textbf{87}, 063625 (2013).

\bibitem{1D sol 4} Y. V. Kartashov, V. V. Konotop, and F. Kh. Abdullaev, Gap
Solitons in a Spin-Orbit-Coupled Bose-Einstein Condensate, \prl\textbf{111},
060402 (2013).

\bibitem{Cardoso} L. Salasnich, W. B. Cardoso, and B. A. Malomed, Localized
modes in quasi-two-dimensional Bose-Einstein condensates with spin-orbit and
Rabi couplings, \pra\textbf{90}, 033629 (2014).

\bibitem{Lobanov} V. E. Lobanov, Y. V. Kartashov, and V. V. Konotop,
Fundamental, Multipole, and Half-Vortex Gap Solitons in Spin-Orbit Coupled
Bose-Einstein Condensates, \prl\textbf{112}, 180403 (2014).

\bibitem{2D SOC gap sol Raymond} Y. Li, Y. Liu, Z. Fan, W. Pang, S. Fu, and
B. A. Malomed, Two-dimensional dipolar gap solitons in free space with
spin-orbit coupling, \pra\textbf{95}, 063613 (2017).

\bibitem{SOC 2D gap sol Hidetsugu} H. Sakaguchi and B. A, Malomed, One- and
two-dimensional gap solitons in spin-orbit-coupled systems with Zeeman
splitting, \pra\textbf{97}, 013607 (2018).

\bibitem{low-dim SOC} Y. V. Kartashov, L. Torner, M. Modugno, E. Ya.
Sherman, B. A. Malomed, and V. V. Konotop, Multidimensional hybrid
Bose-Einstein condensates stabilized by lower-dimensional spin-orbit
coupling, Phys. Rev. Research \textbf{2}, 013036 (2020).

\bibitem{Subhasis}  S. Sinha, R. Nath, and L. Santos,
Trapped Two-Dimensional Condensates with Synthetic Spin-Orbit Coupling, 
\prl\textbf{107}, 270401 (2011).


\bibitem{pra-033621} S. Rojas-Rojas, R. A. Vicencio, M. I. Molina, F. K. Abdullaev, Nonlinear localized modes in dipolar Bose-Einstein condensates in optical lattices, \pra \textbf{84}, 
033621 (2011).

\bibitem{prl-115-253902} Y.-C. Zhang, Z.-W. Zhou, B. A. Malomed, and H. Pu, Stablesolitons in three dimensional free space without the ground state: Self-trapped Bose-Einstein condensates with spin-orbit coupling, \prl\textbf{115}, 253902 (2015).

\bibitem{pra-053608} H.-Y. Chen, Y. Liu, Q. Zhang, Y.-H. Shi, W. Pang, Y.-Y. Li, Dipolar matter-wave solitons in two-dimensional anisotropic discrete lattices, \pra \textbf{93}, 053608 (2016).

\bibitem{pla-418-127696} S. Sabari, R. TamilThiruvalluvar, R. Radha, Modulational instability of spin-orbit coupled Bose-Einstein condensates in discrete media, Phy. Lett. A \textbf{418}, 127696 (2021).

\bibitem{pre-103-022204} J.-C. Liang, Y.-C. Zhang, C. Jiao, A.-X. Zhang,  J.-K. Xue, Ground-state phase and superfluidity of tunable spin-orbit-coupled Bose-Einstein condensates, \pre \textbf{103}, 022204 (2021).

\bibitem{fop-17-6} C. Jiao, J.-C. Liang, Z.-F. Yu, Y. Chen, A.-X. Zhang, J.-K. Xue, Bose-Einstein condensates with tunable spin-orbit coupling in the two-dimensional harmonic potential: The ground-state phases, stability phase diagram and collapse dynamics, Front. Phys. \textbf{17}, 6 (2022).

\bibitem{pra-109-053307} J.-C. Liang, A.-X. Zhang, J.-K. Xue , Ground-state phases and collective modes of supersolid spin-orbit-coupled Bose-Einstein condensates, \pra \textbf{109}, 053307 (2024).

\bibitem{Spielman} I. B. Spielman, Light induced gauge fields for ultracold
neutral atoms, Annual Rev. Cold At. Mol. \textbf{1}, 145 (2012).

\bibitem{Galitski} V. Galitski and I. B. Spielman, Spin-orbit coupling in
quantum gases, \nat\textbf{494}, 49-54 (2013).

\bibitem{Ohberg} N. Goldman, G. Juzeli\={u}nas, P. \"{O}hberg, and I. B.
Spielman, Light-induced gauge fields for ultracold atoms, Rep. Prog. Phys.
\textbf{77}, 126401 (2014).

\bibitem{Zhai} H. Zhai, Degenerate quantum gases with spin-orbit coupling: a
review, Rep. Prog. Phys. \textbf{78}, 026001 (2015).

\bibitem{SOC-sol-review} B. A. Malomed, Creating solitons by means of
spin-orbit coupling, EPL \textbf{122}, 36001 (2018).

\bibitem{PhysRevA.108.044210} B. Liu, X.-Y. Cai, X.-Z. Qin, X.-D. Jiang, J.-N. Xie, B. A. Malomed, Y.-Y. Li, Ring-shaped quantum droplets with hidden vorticity in a radially periodic potential, \pre\textbf{108}, 044210 (2024).

\bibitem{pra-011607} T. Kawakami, T. Mizushima, K. Machida,,Textures of F=2 spinor Bose-Einstein condensates with spin-orbit coupling, \pra \textbf{84}, 011607 (2011).

\bibitem{pra-023606} B. Ramachandhran, B. Opanchuk, X.-J. Liu, H. Pu, P. D. Drummond, H. Hu,  Half-quantum vortex state in a spin-orbit-coupled Bose-Einstein condensate, \pra \textbf{85}, 023606 (2012).

\bibitem{prl-015301} T. Kawakami, T. Mizushima, M. Nitta, K. Machida, Stable Skyrmions in SU(2) Gauged Bose-Einstein Condensates, \prl \textbf{109}, 015301 (2012).

\bibitem{PhysRevA.106.063311} H.-B. Luo, B. A. Malomed, W.-M. Liu, and L.
Li, Tunable energy-level inversion in spin-orbit-coupled Bose-Einstein
condensates, \pra\textbf{106}, 063311 (2022).

\bibitem{PhysRevA.109.013326} H.-B. Luo, L. Li, B. A. Malomed, Y. Li, and B.
Liu, Energy-level inversion for vortex states in spin-orbit-coupled
Bose-Einstein condensates, \pra\textbf{109}, 013326 (2024).

\bibitem{duan} H. Duan,  L. You, B. Gao, Ultracold collisions in the presence of synthetic spin-orbit coupling, \pra\textbf{87}, 052708 (2013).  

\bibitem{Cui} X.-L. Cui, T.-L. Ho, Spin-orbit-coupled one-dimensional Fermi gases with infinite repulsion, \pra\textbf{89}, 013629 (2014).

\bibitem{wang} S.-J. Wang, C.-H. Greene, General formalism for ultracold scattering with isotropic spin-orbit coupling, \pra\textbf{91}, 022706 (2015).

\bibitem{Ma-Jia} D. Ma and C. Jia, Soliton oscillation driven by spin-orbit
coupling in spinor condensates, \pra \textbf{100}, 023629 (2019).

\bibitem{Adhikari} S. K. Adhikari, Phase separation of vector solitons in
spin-orbit-coupled spin-1 condensates, \pra \textbf{100}, 063618 (2019).

\bibitem{pra-023629} 
D.-C. Ma, C.-L. Jia, Soliton oscillation driven by spin-orbit coupling in spinor condensates, \pra \textbf{100}, 023629 (2019).

\bibitem{pra-023617} B. J. Dabrowska-Wüster, E. A. Ostrovskaya, T. J. Alexander, Y. S. Kivshar, Multicomponent gap solitons in spinor Bose-Einstein condensates, \pra \textbf{75}, 023617 (2007).

\bibitem{ho} T.-L. Ho, Spinor Bose Condensates in Optical Traps, \prl\textbf{81}, 4 (1998).

\bibitem{cpb-030302} J. Li, T.-C. He, J. Bai, B. Liu, H.-Y. Wang, Spin-orbit-coupled spin-1 Bose-Einstein condensates confined in radially periodic potential, Chin. Phys. B, \textbf{30}, 030302 (2021).

\bibitem{pra-061601} M. Tsubota, Y. Aoki, K. Fujimoto, Spin-glass-like behavior in the spin turbulence of spinor Bose-Einstein condensates, \pra \textbf{88}, 061601 (2013).

\bibitem{pra-023618} M. Takahashi, V. Pietilä, M. Möttönen, T. Mizushima, K. Machida, Vortex-splitting and phase-separating instabilities of coreless vortices in F=1 spinor Bose-Einstein condensates, \pra \textbf{79}, 023618 (2009).


\bibitem{pra-023641} Q. Guan, D. Blume, Spin structure of harmonically trapped one-dimensional atoms with spin-orbit coupling, \pra \textbf{92}, 023641 (2015).

\bibitem{BEC-SOC GP eqns} Y. Zhang, L. Mao, and C. Zhang, Mean-field
dynamics of spin-orbit coupled Bose-Einstein condensates, \prl\textbf{108},
035302 (2012).

\bibitem{njop-063037} Y.-K. Liu, S.-J. Yang, Stable double-pair skyrmion in an antiferromagnetic F=1 Bose-Einstein condensate, New J. Phys. \textbf{19}, 063037 (2017).


\bibitem{prl-035302} Y.-P. Zhang, L. Mao, C.-W. Zhang, Mean-Field Dynamics of Spin-Orbit Coupled Bose-Einstein Condensates, \prl \textbf{108}, 035302 (2012).
	
\bibitem{pra-043019} J. Zamanian, M. Marklund, G. Brodin, Scalar quantum kinetic theory for spin-1/2 particles: mean field theory, \pra \textbf{12}, 043019 (2010).
	
\bibitem{landau} L. D. Landau and E. M. Lifshitz, \textit{Quantum Mechanics: Non-relativistic Theory} (Nauka Publishers, Moscow, 1974).

\bibitem{pra-033319} K. E. Wilson, E. C. Samson, Z. L. Newman, B. P. Anderson, Generation of high-winding-number superfluid circulation in Bose-Einstein condensates, \pra \textbf{106}, 
033319 (2022).


\bibitem{pra-023612} N. Argaman, Y. B. Band, Finite-temperature density-functional theory of Bose-Einstein condensates, \pra \textbf{83}, 023612 (2011).



\bibitem{chiofalo} M. L. Chiofalo, S. Succi, and M. P. Tosi, Ground state of
trapped interacting Bose-Einstein condensates by an explicit imaginary-time
algorithm, \pre \textbf{62}, 7438 (2000).

\bibitem{Bao} W. Z. Bao and Q. Du, Computing the ground state solution of
Bose-Einstein condensates by a normalized gradient flow, SIAM J. Sci. Comp.
\textbf{25}, 1674 (2004).

\bibitem{njop-043019} A. Saboo, S. Halder, M. Thudiyangal, S. Majumder, Magnetization induced skyrmion dynamics of a spin-orbit-coupled spinor condensate under sinusoidally varying magnetic field, New J. Phys. \textbf{27}, 043019 (2025).


\bibitem{prb-094420} X.-C. Zhang, J. Xia, Y. Zhou, D.-W. Wang, X.-X. Liu, W.-S. Zhao, M. Ezawa, Control and manipulation of a magnetic skyrmionium in nanostructures, \prb \textbf{94}, 094420 (2016).

\end{thebibliography}
\end{document}